\RequirePackage{lineno}
\documentclass[preprint,prd,tightenlines]{revtex4}

\usepackage{graphicx} 
\usepackage{dcolumn}  
\usepackage{colordvi}
\usepackage{color}
\usepackage{epstopdf}
\usepackage{amssymb}
\usepackage{url}
\graphicspath{{ps}}
\usepackage{hyperref}
\usepackage{tabularx}
\usepackage{amsmath}
\usepackage{booktabs}
\usepackage[toc,page]{appendix}
\usepackage{placeins}
\usepackage{subfigure}
\usepackage{caption}
\usepackage{url}
\usepackage{hyperref}

\graphicspath{{./figures/}}

\renewcommand{\thesection}{\arabic{section}}
\renewcommand{\thesubsection}{\thesection.\arabic{subsection}}

\newcommand{\labelsubseccounter}[1]{
    \renewcommand\thesubsection{\kern-0.1em.\arabic{subsection}}
    \addtocounter{subsection}{-1}
    \refstepcounter{subsection}
    \label{#1}
    \renewcommand\thesubsection{\thesection.\arabic{subsection}}
}

\newcommand{\stat}{\ensuremath{\hbox{(stat.)}}}
\newcommand{\syst}{\ensuremath{\hbox{(syst.)}}}

\include{belle2sym}

\begin{document}

\vspace*{-3\baselineskip}
\resizebox{!}{3cm}{\includegraphics{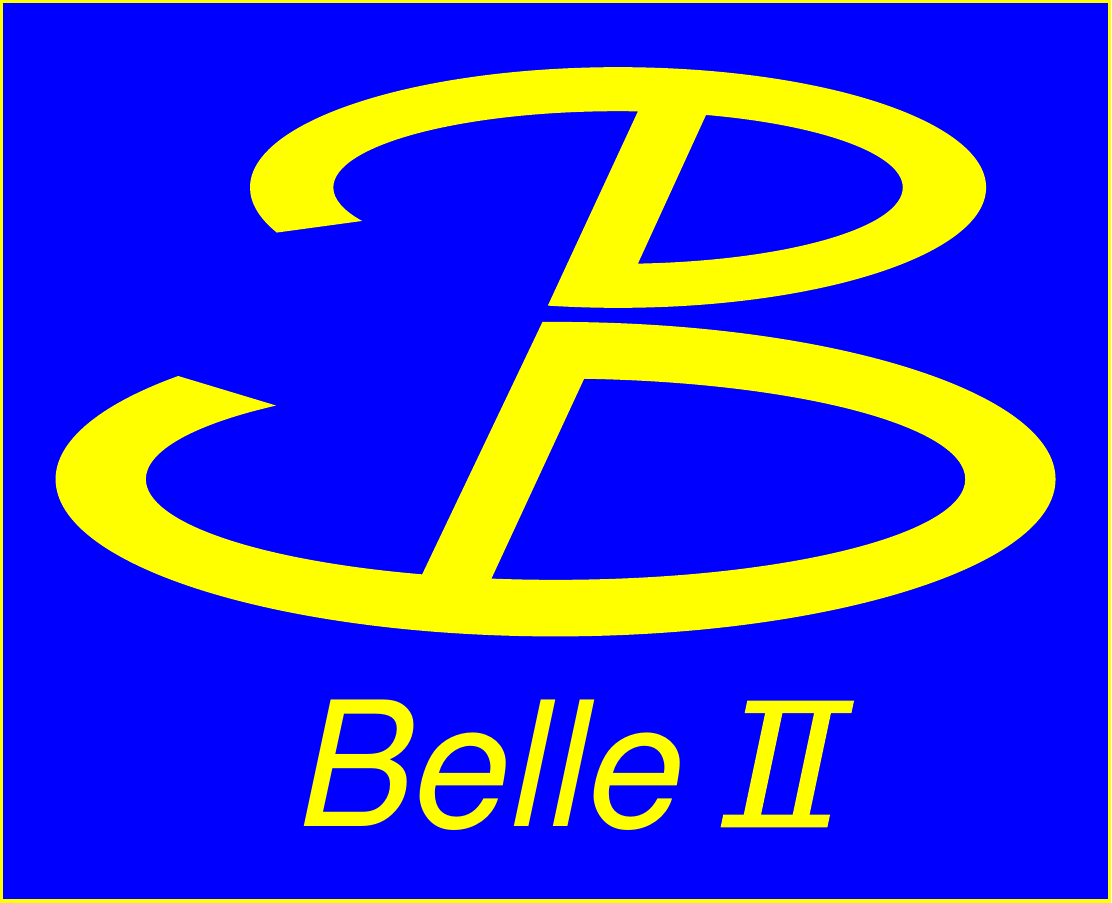}}

\vspace*{-5\baselineskip}
\begin{flushright}
BELLE2-CONF-2022-008\\
% \today\\
% Version 2.31\\
\end{flushright}

\title { \quad\\[0.5cm] Measurement of the branching fraction of the \boldmath{\Bkspzg} decay using \boldmath{$190$~\fbi} of Belle~II data}

\author{Belle II Collaboration: F. Abudin{\'e}n, I. Adachi, K. Adamczyk, L. Aggarwal, P. Ahlburg, H. Ahmed, J. K. Ahn, H. Aihara, N. Akopov, A. Aloisio
% ...
}
\noaffiliation
% \linenumbers

\begin{abstract}
	We report the measurement of the branching fraction of the \Bkspzg decay in $\epem \to \Upsilon(4S) \to B \bar{B}$ data recorded by the Belle~II experiment at the \mbox{SuperKEKB} asymmetric-energy collider and corresponding to $190$~\fbi of integrated luminosity. The signal yield is measured to be $121\pm 29\,\stat$, leading to the branching fraction ${\cal B}\left(\Bkspzg\right) = \left(7.3 \pm 1.8\,\stat \pm 1.0\,\syst \right)\times10^{-6}$, which agrees with the known value.
\end{abstract}


\maketitle

{\renewcommand{\thefootnote}{\fnsymbol{footnote}}}
\setcounter{footnote}{0}

\newpage

%%%%%%%%% INTRO %%%%%%%%%%%%%%%%%%%%%%%%%%%%%%%%%%%%%%%%%%%%%%%%%%%%%%%%%%%%%%%%%%%%%%%%%%%%%%%%%%%%%%%%%%%%%%%%%%%%%%%%%%%%%%%%%%%%%%%%%%%%%%%%%%%%%%%%%%%%%%%%%

\section{Introduction}
\label{sec:introduction}

In the Standard Model (SM), \btsg transitions are forbidden at tree level and are possible only through a quantum loop. Due to the chiral structure of the SM, the radiated photon in \btsg transitions is predominantly left-handed for $b$ and right-handed for $\bar{b}$ quarks. This makes \BzBzb interference in such decays less probable and leads to a suppression, proportional to the $s$ quark mass over the $b$ quark mass, of the decay-time-dependent {\it CP}-violating asymmetry between \Bz and \Bzb decay rates. A broad class of non-SM physics scenarios~\cite{PhysRevLett:79185} feature a different chiral structure and thus may lead to deviations from the SM expectation by introducing different photon polarizations in the transition. 
These non-SM models can be probed via measurements of time-dependent {\it CP} asymmetry (TDCPV) parameters at $B$-Factories.

The \btsg\ process with the highest branching fraction is the \Bkspzg decay, which is produced through the $K^{*0}(892)$ resonance and higher-mass kaonic resonances, such as $K_1^0(1270)$ and $K^{*0}(1410)$, often denoted as \Xsd. Previous TDCPV measurements with this channel have been reported by the Belle and BaBar collaborations~\cite{Ushiroda:2006fi, Aubert:2008gy}. They do not depart from the SM prediction. However their precision is still limited by the small signal sample sizes, which motivates further exploration of the \btsg transitions using the large data set expected from the Belle II experiment at the SuperKEKB collider.

The latest Belle~II branching fraction measurement of $B\to K^* \gamma$ \cite{BelleII:2021tzi} was based on 63~\fbi and was optimised to measure the {\it CP}-violating parameter $A_{\it CP}$ and the isospin
violating-parameter $\Delta_{0+}$ when Belle~II will accumulate more data. In this paper, we report the measurement of the branching fraction of the \Bkspzg decay channel (and its charge conjugate) with decays restricted to those with \KS\piz mass smaller than $1.1\gevcc$ corresponding to the region dominated by the \Kstarz resonance, as a forerunner of the full TDCPV analysis. The analysis uses 190~\invfb of data, corresponding to the luminosity integrated by Belle~II available by winter 2021.

The outline of this paper is as follows. A description of the Belle~II detector and the data set used is given in Section \ref{sec:datasets}. In Section \ref{sec:procedure}, the candidate reconstruction, the selection and the yield extraction method are explained. The result of the measurement of the branching fraction is given in Section \ref{sec:results} and all source of systematic uncertainties are detailed. Finally, the result is summarized in Section \ref{sec:conclusion}. 

%%%%%%%%% BELLE II & data set %%%%%%%%%%%%%%%%%%%%%%%%%%%%%%%%%%%%%%%%%%%%%%%%%%%%%%%%%%%%%%%%%%%%%%%%%%%%%%%%%%%%%%%%%%%%%%%%%%%%%%%%%%%%%%%%%%%%%%%%%%%%%%%%%%%%%%%%%%%%%%%

\FloatBarrier

\section{The Belle~II detector and data set}
\label{sec:datasets}

The Belle II experiment~\cite{Abe:2010sj} operates at the SuperKEKB asymmetric-energy electron-positron collider~\cite{AKAI2018188}, located at the KEK laboratory in Tsukuba, Japan. It consists of various subsystems.
The innermost subsystem is the vertex detector, which includes two layers of silicon pixel detectors and four outer layers of silicon strip detectors. Currently, the second pixel layer covers approximately only 15\% of the azimuthal range, while the remaining vertex detector layers are fully installed. Most of the tracking volume is occupied by a helium and ethane-based small-cell drift chamber (CDC). Outside the drift chamber, a Cherenkov-light imaging and time-of-propagation detector provides charged-particle identification in the barrel region. In the forward endcap, this function is provided by a proximity-focusing, ring-imaging Cherenkov detector with an aerogel radiator. Further out is the electromagnetic calorimeter (ECL), consisting of a barrel and two endcap sections made of CsI(Tl) crystals. A uniform 1.5 T magnetic field is provided by a superconducting solenoid situated outside the calorimeter. Multiple layers of scintillators and resistive plate chambers, located between the magnetic flux-return iron plates, constitute the \KL and muon identification system. The $z$ axis of the laboratory frame is defined as the symmetry axis of the solenoid, pointing approximately at the incoming electron beam.\\

The data used in this analysis were collected between March 2019 and July 2021. They correspond to a total integrated luminosity of $190~\fbi$ obtained at the center-of-mass energies at or near the $\Upsilon(4S)$ resonance (10.58 GeV). The sample corresponds to $N_{B\overline{B}} = (197 \pm 6)\times 10^6$ \BzBzb and \BpBm events determined following Ref.~\cite{nbb}.
To optimize the analysis procedure and determine the detection and selection efficiency, two sets of events simulated using Monte Carlo (MC) are used. The first simulation sample, hereafter referred to as the signal MC sample, contains two million \BzBzb events where one of the $B$ is forced to decay in the signal decay channel. A second sample equivalent to an integrated luminosity of 700~\fbi, referred as generic MC, combines all possible event types \ep\en \to \BzBzb, \BpBm, \uubar, \ddbar, \ccbar and \ssbar. The signal and \ep\en \to \FourS \to \BB samples are generated using the EvtGen package~\cite{LANGE2001152}, while the continuum background events are generated with the KKMC~\cite{kkmc} generator interfaced with Pythia~\cite{pythia}; the detector response is then simulated by the Geant4 package~\cite{AGOSTINELLI2003250}.

Both data and simulated data sets are analysed with Belle II analysis software framework, \texttt{basf2}~\cite{Kuhr:2018cb}.

%%%%%%%%%PROCEDURE%%%%%%%%%%%%%%%%%%%%%%%%%%%%%%%%%%%%%%%%%%%%%%%%%%%%%%%%%%%%%%%%%%%%%%%%%%%%%%%%%%%%%%%%%%%%%%%%%%%%%%%%%%%%%%%%%%%%%%%%%%%%%%%%%%%%%%%%%%%%%%%

\FloatBarrier
\section{Analysis}
\label{sec:procedure}

%%%%%%%%%RECO+SELECTION%%%%%%%%%%%%%%%%%%%%%%%%%%%%%%%
\subsection{Candidate reconstruction and selection}
\label{sec:recoselec}
\labelsubseccounter{subsec:recoselec}

We search for candidate \Bz decays with five or more charged particles in the event, at least one energy deposit (cluster) in the ECL larger than 0.2~\gev, and at least 4~\gev of visible energy in the center-of-mass frame to enrich the sample in events with high-energy photons. We reconstruct \KS candidates from the association of two oppositely charged particles within the CDC acceptance and originating from the interaction point. For these particles, the distance of closest approach to the \epem interaction point is required to be smaller than 2.0 cm in the plane transverse to the $z$ axis and smaller than 4.0~cm along the $z$ axis. The \KS properties are obtained from a kinematic fit of the trajectories of both charged particles (tracks) assumed to be pions. Events with \KS candidates with mass outside the range 0.450 to 0.550 \gevcc are discarded.
We reconstruct \piz candidates from the combination of two photons, each having an energy of at least 30~\mev in the barrel, 80~\mev in the forward endcap or 60~\mev in the backward endcap of the ECL. Only pairs with an invariant mass within 0.120 to 0.145 \gevcc are kept.
High-energy photons are selected if their energy in the laboratory frame is comprised between 1.4 and 4.0~\gev. With a \piz and $\eta$ veto, consisting in a boosted decision-tree (BDT) trained on event-based variables, we reject photons consistent with products of \piz and $\eta$ diphoton decays with 70\% probability or more.

Finally, the reconstructed \Bz candidates combine one \KS, one \piz and one high-energy photon in a  kinematic fit including a pointing constraint to the interaction point~\cite{treefitter}. Then we construct two variables based on the center-of-mass energy of the collisions, $\sqrt{s}$, the $B$ invariant mass constrained by the beam energy \mbc = $\sqrt{(\sqrt{s}/2)^{2} - p^{*2}_{B}}$ and the energy difference \de $= E^{*}_{B} - \sqrt{s}/2$, where $p^{*}_{B}$ and $E^{*}_{B}$ are respectively the \Bz candidate momentum and energy in the center-of-mass frame. We require that $5.20 < \mbc < 5.29$ \gevcc and $ -0.5 < \de < 0.5$ \gev.

This first reconstruction and selection step retains approximately 28$\%$ of the signal decay. The signal significance, $S/\sqrt{S+B}$ where $S$ and $B$ are respectively the number of signal decays and background events from simulation, is still too low and further selection requirements are needed. Additional variables are considered: a threshold of 36 for the \KS decay length significance, computed from the ratio of the decay length over its uncertainty, is used to suppress background in the \KS sample; a maximal value of $1.1\gevcc$ for the $M(\KS\piz)$ mass allows to retain most of the \Kstarz decays and reject the ones from the $X_{sd}$ resonances; a threshold of $5.275~\gevcc$ is required on the beam-constrained mass \mbc; finally, to suppress the dominant background originating from continuum events \epem \to \qqbar, we combine thirty event-shape variables and kinematic properties of the event into a binary BDT classifier \cite{foxwolfram} trained over 60000 signal events and an equal amount of simulated background events.
These four selection criteria values are optimized to maximize the significance over the 700~\fbi generic MC data sample. 
After this selection, for events containing multiple candidates, the one featuring the lowest $\chi^2$ from the \Bz vertex fit is chosen and all others discarded. There is only one candidate on more than half of the event and the efficiency to select the correctly reconstructed signal from an event with multiple reconstructed $B$ candidates is 87\%.

The efficiency is then corrected from the discrepancies observed between data and simulation described in Section \ref{subsec:systematics}. The final total selection efficiency reaches $\epsilon = (8.7 \pm 0.7)\%$, where the quoted uncertainty comes from the total systematic uncertainty from Table \ref{table:syst}.\\

%%%%%%%%%FIT%%%%%%%%%%%%%%%%%%%%%%%%%%%%%%%
\FloatBarrier
\subsection{Yield extraction}
\label{sec:sigextract}
\labelsubseccounter{subsec:sigextract}

\begin{figure}
\centering
\begin{subfigure}
  \centering
  \includegraphics[width=0.49\linewidth]{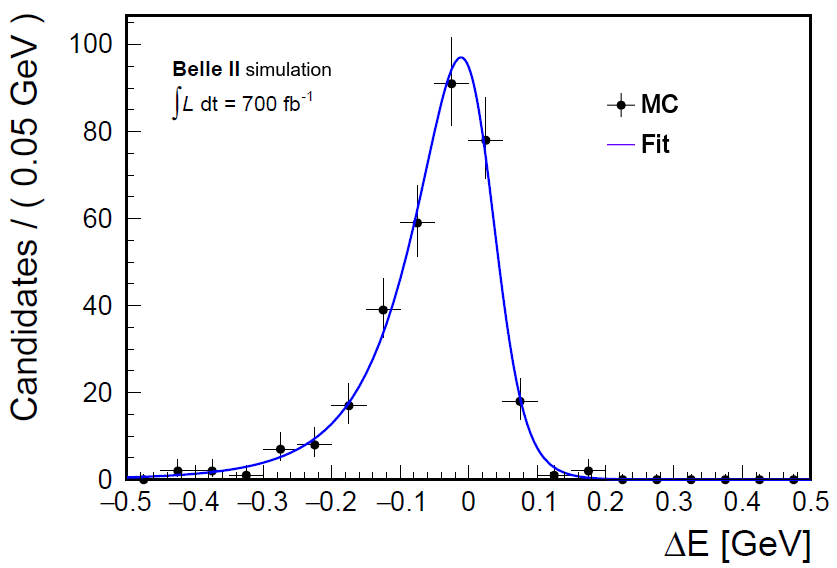}
\end{subfigure}
\begin{subfigure}
  \centering
  \includegraphics[width=0.49\linewidth]{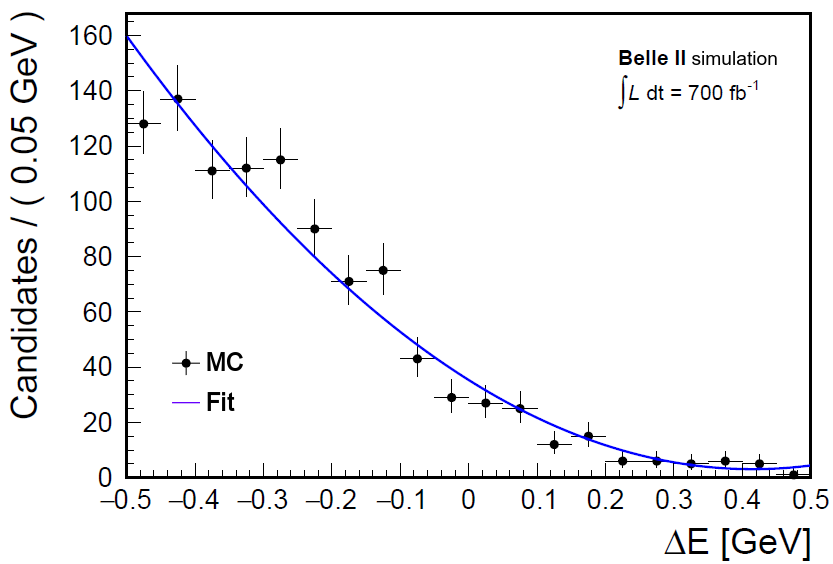}
\end{subfigure}
\caption{\sl Distribution of \de for candidates selected in simulated samples corresponding to 700~\fbi of (left) signal \Bkspzg decays and (right) \epem \to \FourS events with modeling-fit projections overlaid.}
\label{fig:fitshapeskspzg}
\end{figure}

The signal yield is extracted from an extended maximum likelihood fit to the \de distribution, where the total probability density function includes two contributions, one for the signal and the other for the background. A single Chebyshev polynomial of order two is used to model the background contribution. The two parameters $a_0$ and $a_1$ are fixed from a fit to the 700~\fbi generic MC sample, shown in Figure \ref{fig:fitshapeskspzg}.
A Johnson distribution \cite{johnson}, which acts like a double-sided Crystal Ball function~\cite{cb}, describes the signal contribution with four parameters, one associated with the location of the peak, and three with the shape. The parametrization of the Johnson distribution reads

\begin{equation}
\label{eq:johnson}
\mathrm{PDF}_{\mathrm{Johnson}}(\de; \mu, \sigma, \lambda, \gamma) = \frac{\lambda}{\sigma\sqrt{2\pi}} \frac{1}{\sqrt{1 + \left( \frac{\de-\mu}{\sigma} \right)^2}} \;\exp\left[-\frac{1}{2} \left(\gamma + \lambda\, \text{arsinh} \left(\frac{\de-\mu}{\sigma}\right) \right)^2\right],
\end{equation}
where $\mu$ and $\sigma$, which drive the central value and width of the distribution, are determined by the fit and the parameters $\lambda$ and $\gamma$ are fixed from a fit to the signal MC sample (see Figure \ref{fig:fitshapeskspzg}).
The third (and last) floating parameter in the fit is the signal yield.

%%%%%%%%%RESULTS and SYSTEMATICS%%%%%%%%%%%%%%%%%%%%%%%%%%%%%%%%%%%%%%%%%%%
%%%%%%%%%%%%%%%%%%%%%%%%%%%%%%%%%%%%%%%%%%%%%%%%%%%%%%%%%%%%%%%%%%%%%%%%%%%

\FloatBarrier
\section{Results and systematic uncertainties}
\label{sec:results}

%%%%%%%%%BRANCHING FRACTION%%%%%%%%%%%%%%%%%%%%%%%%%%%%%%%%%%%%%%%%%%%%%%%

\subsection{Measurements of the branching fraction}
\label{sec:measuredBR}

\begin{figure}[h!]
	{\centering
		\includegraphics[width=.6\linewidth]{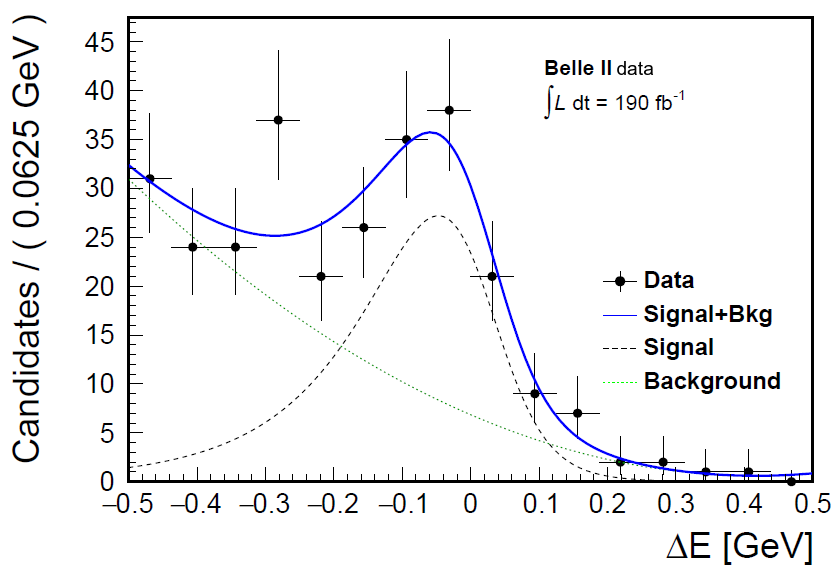}
		\caption{\sl Distribution of \de for candidates selected in data sample corresponding to 190~\fbi of \Bkspzg decays with modeling-fit projections overlaid.}
		\label{fig:finalfitkspzg}
	}
\end{figure}

The fit described in Section \ref{subsec:sigextract} is applied to the \de distribution of the selected candidates. Figure \ref{fig:finalfitkspzg} depicts the data and the fitted function overlaid. 
The observed \Bkspzg yield is $N_{\mathrm{yield}} = 121 \pm 29\,\stat$. 

The branching fraction is computed from\\
\begin{equation}
\label{eq:bfraction}
{\cal B} = \frac {N_{\mathrm{yield}}} {2 \epsilon f^{00} N_{B\overline{B}} },
\end{equation}
where $\epsilon$ is the overall detection and selection efficiency for the \Bkspzg decay estimated in Section \ref{subsec:recoselec}, $f^{00} = (48.6\pm0.6)\%$ \cite{PDG} is the branching fraction of \FourS going to $\BzBzb$, and $N_{B\overline{B}}$ is the quoted number of  ${B\overline{B}}$ pairs produced in the data sample listed in Section \ref{sec:datasets} \cite{nbb}. 
We find the following branching fraction
\begin{equation}
    {\cal B}\left(\Bkspzg\right) = \left(7.3 \pm 1.8\,\stat  \right) \times 10^{-6}.
\end{equation}

%%%%%%%%%SYSTEMATICS%%%%%%%%%%%%%%%%%%%%%%%%%%%%%%%%%%%%%%%%%%%%%%%%%%%%%%

\FloatBarrier
\subsection{Systematic uncertainties}
\label{sec:systematics}
\labelsubseccounter{subsec:systematics}

Systematic uncertainties arising from the generation, reconstruction, selection, fit and yield extraction procedures are considered and discussed below. All relative systematic uncertainty values are summarized in Table~\ref{table:syst}.
A systematic uncertainty of 0.2\% is assigned due to the finite sample size of the MC signal sample.
To assign the systematic uncertainty related to the model of the $K^{*0}$ resonance, we check the efficiencies of the $M(\KS\pi^0)$ restriction after the preselection and reconstruction and after the whole selection except for this criterion. We assign a relative systematic uncertainty of 2.0\% to the efficiency.
For the \piz selection, the difference in reconstruction efficiency between data and simulation is estimated by comparing the $\eta \to \gamma\gamma$ and $\eta \to \piz\piz\piz$ yields. A relative uncertainty of 5.5\% is assigned.
Comparing the reconstruction efficiency of \KS between data and simulation, a systematic uncertainty of 3.5\% and a correction factor of 0.9654 is assigned to the efficiency.
The systematic uncertainty related to the \piz and $\eta$ veto is evaluated by comparing the efficiency of that veto between real and simulated $\Bz \to D^{-}\pi^+$ and $\Bp \to \overline{D}^{0}\pi^+$ samples. The efficiency is extracted as a function of the photon energy. We assign a systematic uncertainty of 1.9\% and a correction factor of 1.034 on the efficiency.
According to the measurement of the data to MC ratio of photon reconstruction efficiency in the calorimeter, using radiative muon pair events, we assign a relative uncertainty of 0.3\% for the photon selection.
The systematic uncertainty due to the discrepancies between data and simulation of the efficiency associated with the BDT selection in the off-resonance data, recorded 60~\mev below \FourS resonance, is evaluated to be 3.0\%. 
To check the estimator properties, we perform a MC study: 500 data sets, each corresponding to an integrated luminosity of 200~\fbi, are created by bootstrapping the initial 700~\fbi MC sample. 
Then the final fit procedure is performed for each data set. 
We observe an overestimation in the signal yield. We tried to reduce this bias by testing different fitting function or fitting the peaking \BpBm background around $-0.3~\gev$, but no improvements were observed. We assign a systematic uncertainty of 11.5\%.
A systematic uncertainty of 2.9\% is assigned due to the uncertainty in the number of produced $B\bar{B}$ pairs. A systematic uncertainty of 1.2\% is assigned due to the uncertainty in the branching fraction of \FourS decaying to $B\overline{B}$ from Ref.~\cite{PDG}.

\begin{table}[]
\caption{\sl Summary of systematic uncertainties considered for the measurement of the \Bkspzg branching fraction. The total efficiency systematic corresponds to the quadratic sum of all the systematic uncertainties for the selection efficiency.}
\begin{tabular}{lr}
\hline
\multicolumn{1}{|l|}{MC sample size}          &  \multicolumn{1}{c|}{0.2\%}        \\ 
\multicolumn{1}{|l|}{MC generation}          &  \multicolumn{1}{c|}{2.0\%}        \\ 
\multicolumn{1}{|l|}{\piz reconstruction}    &  \multicolumn{1}{c|}{5.5\%}    \\ 
\multicolumn{1}{|l|}{\KS reconstruction}     &  \multicolumn{1}{c|}{3.5\%} \\ 
\multicolumn{1}{|l|}{\piz-$\eta$ veto}              &  \multicolumn{1}{c|}{1.9\%}        \\ 
\multicolumn{1}{|l|}{$\gamma$ selection}     & \multicolumn{1}{c|}{0.3\%}          \\ 
\multicolumn{1}{|l|}{Continuum suppression}     & \multicolumn{1}{c|}{3.0\%}          \\  \hline
\multicolumn{1}{|l|}{Total efficiency}     &  \multicolumn{1}{c|}{7.7\%} \\ \hline \hline
\multicolumn{1}{|l|}{Fit bias}                 & \multicolumn{1}{c|}{11.5\%}        \\ \hline \hline
\multicolumn{1}{|l|}{Number of \BzBzb pairs}   & \multicolumn{1}{c|}{2.9\%}       \\ \hline \hline
\multicolumn{1}{|l|}{$f^{00}$ systematic}   & \multicolumn{1}{c|}{1.2\%}       \\ \hline \hline
\multicolumn{1}{|l|}{Total systematic on ${\cal B}$}   & \multicolumn{1}{c|}{14.2\%}        \\
\hline
\end{tabular}
\label{table:syst}
\end{table}

%%%%%%%%%CONCLUSION%%%%%%%%%%%%%%%%%%%%%%%%%%%%%%%%%%%%%%%%%%%%%%%%%%%%%%%%%%%%%%%%%%%%%%%%%%%%%%%%%%%%%%%%%%%%%%%%%%%%%%%%%%%%%%%%%%%%%%%%%%%%%%%%%%%%%%%%%%%%

\FloatBarrier
\section{Conclusion}
\label{sec:conclusion}

Using a sample of data corresponding to $190~\fbi$ recorded with the Belle~II experiment, we report a measurement of the branching fraction for the \Bkspzg decay with a larger sample size than reported in Ref.~\cite{BelleII:2021tzi}. The measured branching fraction for this decay is 
\begin{equation}
    {\cal B}\left(\Bkspzg\right) = \left(7.3 \pm 1.8\,\stat \pm 1.0\,\syst \right) \times 10^{-6},
\end{equation}
which is compatible with the known value of $(7.0 \pm 0.4) \times 10^{-6}$~\cite{PDG}.

\FloatBarrier
\section{Acknowledgments}

We thank the SuperKEKB group for the excellent operation of the
accelerator; the KEK cryogenics group for the efficient
operation of the solenoid and the KEK computer group for
on-site computing support.

\bibliography{belle2}

\providecommand{\href}[2]{#2}\begingroup\raggedright\begin{thebibliography}{10}

\bibitem{PhysRevLett:79185}
D.~Atwood, M.~Gronau, and A.~Soni, {\em {Mixing-Induced $\mathit{CP}$
  Asymmetries in Radiative $\mathit{B}$ Decays in and beyond the Standard
  Model}\/},  \href{http://dx.doi.org/10.1103/PhysRevLett.79.185}{Phys. Rev.
  Lett. {\bf 79} (1997)  185--188},
\href{http://arxiv.org/abs/9704272}{{\tt arXiv:9704272 [hep-ph]}}.
%%CITATION = ARXIV:9704272;%%.

\bibitem{Ushiroda:2006fi}
Y.~Ushiroda et al., {Belle collaboration}, {\em {Time-Dependent CP Asymmetries
  in $B^0 \to K^0_S \pi^0 \gamma$ transitions}\/},
  \href{http://dx.doi.org/10.1103/PhysRevD.74.111104}{Phys. Rev. {\bf D74}
  (2006)  111104},
\href{http://arxiv.org/abs/hep-ex/0608017}{{\tt arXiv:hep-ex/0608017
  [hep-ex]}}.
%%CITATION = HEP-EX/0608017;%%.

\bibitem{Aubert:2008gy}
B.~Aubert et al., {BaBar collaboration}, {\em {Measurement of Time-Dependent CP
  Asymmetry in $B^0 \to K^0_{S} \pi^0 \gamma$ Decays}\/},
  \href{http://dx.doi.org/10.1103/PhysRevD.78.071102}{Phys. Rev. {\bf D78}
  (2008)  071102},
\href{http://arxiv.org/abs/0807.3103}{{\tt arXiv:0807.3103 [hep-ex]}}.
%%CITATION = ARXIV:0807.3103;%%.

\bibitem{BelleII:2021tzi}
F.~Abudin\'en et al., {Belle II collaboration}, {\em {Measurements of the
  branching fractions for $B \to K^{*}\gamma$ decays at Belle II}\/},
  \href{http://arxiv.org/abs/2110.08219}{{\tt arXiv:2110.08219 [hep-ex]}}.

\bibitem{Abe:2010sj}
T.~Abe et al., {Belle II collaboration}, {\em {Belle II Technical Design
  Report}\/},
\href{http://arxiv.org/abs/1011.0352}{{\tt arXiv:1011.0352 [physics.ins-det]}}.
%%CITATION = ARXIV:1011.0352;%%.

\bibitem{AKAI2018188}
K.~Akai, K.~Furukawa, and H.~Koiso, {SuperKEKB}, {\em {SuperKEKB Collider}\/},
  \href{http://dx.doi.org/10.1016/j.nima.2018.08.017}{Nucl. Instrum. Meth. A
  {\bf 907} (2018)  188--199}, \href{http://arxiv.org/abs/1809.01958}{{\tt
  arXiv:1809.01958 [physics.acc-ph]}}.

\bibitem{nbb}
C.~Cecchi et al., {Belle II collaboration}, {\em {B counting measurement in
  "Moriond 2022" Belle II dataset}\/},  BELLE2-NOTE-PH-2022-007 (2022)  .
  \url{https://docs.belle2.org/record/2846/}.

\bibitem{LANGE2001152}
D.~J. Lange, {\em {The EvtGen particle decay simulation package}\/},
  \href{http://dx.doi.org/10.1016/S0168-9002(01)00089-4}{Nucl. Instrum. Meth. A
  {\bf 462} (2001)  152--155}.

\bibitem{kkmc}
B.~F.~L. Ward, S.~Jadach, and Z.~Was, {\em {Precision calculation for e+ e-
  ---\ensuremath{>} 2f: The KK MC project}\/},
  \href{http://dx.doi.org/10.1016/S0920-5632(03)80147-0}{Nucl. Phys. B Proc.
  Suppl. {\bf 116} (2003)  73--77},
  \href{http://arxiv.org/abs/hep-ph/0211132}{{\tt arXiv:hep-ph/0211132}}.

\bibitem{pythia}
T.~Sjostrand et al., {\em {A brief introduction to PYTHIA 8.1}\/},
  \href{http://dx.doi.org/10.1016/j.cpc.2008.01.036}{Comp. Phys. Comm. {\bf
  178} (2008)  852--867}.

\bibitem{AGOSTINELLI2003250}
S.~Agostinelli et al., {GEANT4}, {\em {GEANT4--a simulation toolkit}\/},
  \href{http://dx.doi.org/10.1016/S0168-9002(03)01368-8}{Nucl. Instrum. Meth. A
  {\bf 506} (2003)  250--303}.

\bibitem{Kuhr:2018cb}
T.~Kuhr et al., {Belle II collaboration}, {\em {The Belle II Core Software}\/},
   \href{http://dx.doi.org/10.1007/s41781-018-0017-9}{Comput. Softw. Big. Sci.
  {\bf 3} (2018)  },
\href{http://arxiv.org/abs/1809.04299}{{\tt arXiv:1809.04299
  [physics.comp-ph]}}.
%%CITATION = HEP-EX/1809.04299;%%.

\bibitem{treefitter}
J.-F. Krohn et al., {Belle II collaboration}, {\em {Global decay chain vertex
  fitting at Belle II}\/},
  \href{http://dx.doi.org/10.1016/j.nima.2020.164269}{Nucl. Instrum. Meth. {\bf
  A976} (2020)  164269}, \href{http://arxiv.org/abs/1901.11198}{{\tt
  arXiv:1901.11198 [hep-ex]}}.

\bibitem{foxwolfram}
A.~Bevan et al., {\em {The Physics of the B Factories}\/},
  \href{http://dx.doi.org/10.1140/epjc/s10052-014-3026-9}{Eur. Phys. J. {\bf
  C74} (2014)  3026}.

\bibitem{johnson}
N.~L. Johnson, {\em {Systems of Frequency Curves Generated by Methods of
  Translation}\/},  \href{http://dx.doi.org/10.2307/2332539}{Biometrika {\bf
  36} (1949)  149--176}.

\bibitem{cb}
T.~Skwarnicki, {\em {A study of the radiative CASCADE transitions between the
  Upsilon-Prime and Upsilon resonances}}.
\newblock PhD thesis, Cracow, INP, 1986.
\newblock \url{https://inspirehep.net/literature/230779}.

\bibitem{PDG}
P.~Zyla and others. (Particle Data~Group), {\em {Review of Particle
  Physics}\/},  \href{http://dx.doi.org/10.1093/ptep/ptaa104}{Prog. Theor. Exp.
  Phys. {\bf 2020} (2020) no.~8, 071102}. 083C01.

\end{thebibliography}\endgroup
\bibliographystyle{belle2-note}

\end{document}